\documentclass[sigconf,natbib=true]{acmart}
\AtBeginDocument{%
  \providecommand\BibTeX{{%
    Bib\TeX}}}

\usepackage{graphicx}
\usepackage{subfiles}

\usepackage{todonotes}
\usepackage{booktabs}
\usepackage{multirow}
\PassOptionsToPackage{hyphens}{url}\usepackage{hyperref}

\usepackage{dsfont}
\usepackage{amsmath}

\usepackage{caption}
\usepackage{subcaption}
\usepackage{enumitem}

\usepackage{nicematrix}
\usepackage[dvipsnames]{xcolor}

\usepackage{pifont}
\AtBeginDocument{%
  \providecommand\BibTeX{{%
    \normalfont B\kern-0.5em{\scshape i\kern-0.25em b}\kern-0.8em\TeX}}}

\setcopyright{acmcopyright}
\copyrightyear{2026}
\acmYear{2026}
\acmDOI{10.1145/1122445.1122456}

\acmConference[SIGIR '26]{SIGIR '26: International ACM SIGIR Conference on Research and Development in Information Retrieval}{July 20--24, 2026}{Naarm, Australia}
\acmBooktitle{SIGIR '26: ACM SIGIR Conference on Research and Development in Information Retrieval,
  July 20--24, 2026, Naarm, Australia}

\begin{document}

\clubpenalties 3 1001 1002 0
\widowpenalties 3 2001 2002 0

\title[CoverageBench:Evaluating Information Coverage across Tasks and Domains]{CoverageBench: \\Evaluating Information Coverage across Tasks and Domains}

\settopmatter{authorsperrow=4}

\author{Saron Samuel}
\affiliation{%
  \institution{Johns Hopkins University}
  \city{Baltimore}
  \state{MD}
  \country{USA}
}
\email{ssamue21@jhu.edu}

\author{Andrew Yates}
\affiliation{%
  \institution{Johns Hopkins University}
  \city{Baltimore}
  \state{MD}
  \country{USA}
}
\email{andrew.yates@jhu.edu}

\author{Dawn Lawrie}
\affiliation{%
  \institution{Johns Hopkins University}
  \city{Baltimore}
  \state{MD}
  \country{USA}
}
\email{lawrie@jhu.edu}

\author{Ian Soboroff}
\affiliation{%
  \institution{National Institute of Standards and Technology}
  \city{Gaithersburg}
  \state{MD}
  \country{USA}
}
\email{ian.soboroff@nist.gov}

\author{Trevor Adriaanse}
\affiliation{%
  \institution{Johns Hopkins University}
  \city{Baltimore}
  \state{MD}
  \country{USA}
}
\email{tadriaa1@jhu.edu}

\author{Benjamin Van Durme}
\affiliation{%
  \institution{Johns Hopkins University}
  \city{Baltimore}
  \state{MD}
  \country{USA}
}
\email{vandurme@jhu.edu}

\author{Eugene Yang}
\affiliation{%
  \institution{Johns Hopkins University}
  \city{Baltimore}
  \state{MD}
  \country{USA}
}
\email{eugene.yang@jhu.edu}

\renewcommand{\shortauthors}{Samuel et al.}



\begin{abstract}

We wish to measure the information coverage of an ad hoc retrieval algorithm, that is, how much of the range of available relevant information is covered by the search results.
Information coverage is a central aspect for retrieval, especially when the retrieval system is integrated with generative models in a retrieval-augmented generation (RAG) system.
The classic metrics for ad hoc retrieval, precision and recall, reward a system as more and more relevant documents are retrieved. However, since relevance in ad hoc test collections is defined for a document without any relation to other documents that might contain the same information, high recall is sufficient but not necessary to ensure coverage. The same is true for other metrics such as rank-biased precision (RBP), normalized discounted cumulative gain (nDCG), and mean average precision (MAP). Test collections developed around the notion of diversity ranking in web search incorporate multiple aspects that support a concept of coverage in the web domain. In this work, we construct a suite of collections for evaluating information coverage from existing collections. This suite offers researchers a unified testbed spanning multiple genres and tasks. All topics, nuggets, relevance labels, and baseline rankings are released on Hugging Face Datasets\footnote{\textit{https://huggingface.co/datasets/hltcoe/coveragebench}}, along with instructions for accessing the publicly available document collections.

\end{abstract}



\maketitle

\section{Introduction}

Information Retrieval (IR) evaluation has historically focused on the question, \textit{``can systems find relevant documents?''} Standard benchmarks usually provide binary or graded relevance judgments to measure metrics like nDCG, MAP, and RBP~\cite{ndcg, intro_to_ir, rbp_paper}. These relevance-based evaluations have driven decades of progress in retrieval effectiveness, from statistical ranking methods~\cite{bm25_and_beyond_statistical} to modern neural retrievers~\cite{colbert, spladev2}.

However, relevance alone is insufficient for many real-world information needs~\cite{novelty_diversity_IR_eval, beyond_independent_relevance}. Users often seek \emph{comprehensive} understanding of a topic rather than a handful of relevant document~\cite{novelty_diversity_IR_eval}. 
In fact, this concept has been discussed as early as in TREC-6 interactive track, which was in 1996~\cite{DBLP:conf/trec/Over97}. 
A system might retrieve ten highly relevant documents that all discuss the same aspect of the user query, scoring well on traditional metrics while failing to address the full breadth of the information need~\cite{beyond_independent_relevance, use_of_MMR}. 
However, for a while, models have been struggling to retrieve more relevant documents. Therefore, improving relevance has been the primary focus in retrieval model developments. 
However, recent advancements in neural retrieval, such as reasoning-based rerankers~\cite{rank1, rankk}, enable the direct modeling of information coverage. 
Especially in Retrieval-Augmented Generation (RAG) systems, retrieved content directly informs a generated response~\cite{rag_knowledge_intensive_nlp}. A RAG system that retrieves only a narrow set of information with an abundance of redundant documents, no matter how relevant, will produce incomplete or biased answers~\cite{rag_survey, power_of_noise}.

Coverage evaluation addresses this gap by assessing whether a system surfaces diverse, complementary information that collectively satisfies an information need. Rather than asking, \emph{``is this document relevant?''}, coverage evaluation asks, \emph{``does this set of documents contain all the important information?''} This requires identifying discrete information units (i.e., nuggets) essential to answering a query and measuring how completely a retrieved set covers them~\cite{qa_track, deconstructing_nuggets, eval_summary_answer_2_coins, auto_argue}.

Although aspectual recall seems intuitively to be an important metric, there are relatively few test collections built to measure it. The TREC-6 interactive track~\cite{DBLP:conf/trec/Over97} decorated TREC topics with narrative descriptions of desired aspects to be covered, and searchers in the task were challenged to cover the topic as well as they could, but this did not result in test collection for ad hoc search research. The TREC web track diversity ranking tasks~\cite{DBLP:conf/trec/ClarkeCS09,DBLP:conf/trec/ClarkeCSC10,DBLP:conf/trec/ClarkeCSV11} explored subtopic coverage in the context of web search. Later work at NTCIR explored identifying the subtopics themselves~\cite{DBLP:conf/ntcir/Song0SKLSWO11,DBLP:conf/sigir/SakaiDYLZKSI13,DBLP:conf/ntcir/LiuS0DYKOZ14,DBLP:conf/ntcir/YamamotoLZDZMKO16}. 
The TREC NeuCLIR Report Generation Pilot Task~\cite{neuclir_2024} and TREC RAG Track~\cite{pradeep2024initialnuggetevaluationresults, pradeep2024ragnarokreusableragframework, thakur2025supportevaluationtrec2024} have incorporated coverage evaluation, but with limited query sets and narrow domains. Creating new coverage benchmarks from scratch is expensive, requiring annotation of information nuggets for each topic and nugget-level relevance judgments. 
The sparsity of collections may in part be because algorithms need to be sufficiently advanced to distinguish aspects within documents. The time is now ripe for the development of these algorithms.

To support this, we address this challenge by transforming existing collections into coverage evaluation benchmarks to make CoverageBench, a unified suite of coverage evaluation collections.
CoverageBench includes seven datasets. Five are adapted from established TREC tasks: NeuCLIR 2024~\cite{neuclir_2024}, RAG 2024~\cite{pradeep2024initialnuggetevaluationresults, pradeep2024ragnarokreusableragframework, thakur2025supportevaluationtrec2024}, Fair Ranking 2022~\cite{fair_ranking_2022}, CAsT 2020~\cite{cast2020}, and RAGTIME 2025~\cite{ragtime}. We also include two coverage-based evaluation datasets not from TREC, CRUX-MultiNews~\cite{crux}, CRUX-DUC04~\cite{crux}. These collections were chosen over others because of the availability of the document collections.
To establish baseline performance, we evaluate six retrieval configurations with BM25~\cite{pyserini} and Qwen3-8B~\cite{qwen3embedding}, with and without reranking (Rank1-7B~\cite{rank1} and Qwen3-Reranker-8B~\cite{qwen3embedding}).

Our contributions are the following: 
\begin{itemize}
    \item we present a methodology for deriving coverage benchmarks from existing ad hoc retrieval collections
    \item we release CoverageBench, comprising seven coverage-annotated datasets with publicly available augmentations on Hugging Face Datasets
    \item we provide baseline results and evaluation tools that lower the barrier to coverage-oriented retrieval research.
\end{itemize} 
We enable the IR community to explore coverage evaluation at scale, supporting the development of retrieval systems better suited for RAG and other applications where comprehensive information access matters.
\section{Background}


\subsection{Information Coverage}


Information coverage evaluation addresses the gap of finding relevant material and covering the breadth of an information need by decomposing an information need into discrete units of information, commonly called nuggets, and measuring how completely a retrieved set (or a generated response) accounts for them~\cite{qa_track}. The TREC QA track~\cite{qa_track} introduced nugget-based evaluation, where each question was associated with a set of atomic facts (nuggets) that a correct answer should contain. Similarly, the pyramid method~\cite{nenkova-passonneau-2004-evaluating} for summarization evaluation weights content units by how many human reference summaries include them.

In the retrieval setting, coverage shifts the unit of evaluation from individual documents to the set of retrieved results considered collectively. This perspective is closely related to search result diversification. $\alpha$-nDCG~\cite{novelty_diversity_IR_eval} and intent-aware metrics~\cite{intent_based_metrics} penalize redundancy by discounting documents that cover subtopics already represented in higher-ranked results. Subtopic Recall (StRecall)~\cite{beyond_independent_relevance} directly measures the fraction of distinct subtopics covered by the top-k retrieved documents. 

In a RAG pipeline, the retrieved documents serve as the knowledge context from which a language model generates its response. If the retrieval stage fails to surface a nugget, the generation stage has no opportunity to include that information. Coverage failures at retrieval time can propagate directly into incomplete or biased generated answers~\cite{rag_survey, power_of_noise}. This makes coverage evaluation essential not only as a retrieval quality measure but as a diagnostic tool for understanding end-to-end RAG system behavior.

\subsection{Benchmarks for Information Coverage}



One of the early efforts to annotate a collection for information coverage was  
the 2009-2011 TREC web tracks~\cite{DBLP:conf/trec/ClarkeCS09,DBLP:conf/trec/ClarkeCSC10,DBLP:conf/trec/ClarkeCSV11}, which featured a diversity ranking task. Diversity ranking is a term from web search where the search engine attempts to cover as many aspects of an underspecified query as possible, high up in the result page. As the web track organizers write, ``\textit{The goal of this
diversity task is to return a ranked list of pages that together provide complete coverage for a query, while avoiding excessive redundancy in the result list.}''~\cite{DBLP:conf/trec/ClarkeCSV11}
The topics were derived from the logs of a commercial search engine, using a tool designed to cluster queries with similar click patterns. Topics were divided into two classes: faceted and ambiguous. A faceted topic like ``adobe indian houses'' had subtopics about adobe houses, their origins, how they are built, and how native Americans built them. Each subtopic of a faceted query represented a related information need that might have been meant by the query. An ambiguous query like ``east ridge high school'' had subtopics with orthogonal interpretations: there are East Ridge High Schools in many places in the United States, and the user seemed likely to want only one of them (since ``school'' is singular). Queries of acronyms were also typical, ambiguous queries. Subtopics were identified when the topics were developed, without later expansion. The TREC web track pioneered the intent-aware ERR (IA-ERR) and $\alpha$-nDCG metrics specifically designed to take subtopics into account~\cite{DBLP:conf/wsdm/ClarkeCSA11}.


The INTENT, INTENT2, IMINE, IMINE-2 tasks at NTCIR added an explicit subtopic discovery task to diversity ranking~\cite{DBLP:conf/ntcir/Song0SKLSWO11,DBLP:conf/sigir/SakaiDYLZKSI13,DBLP:conf/ntcir/LiuS0DYKOZ14,DBLP:conf/ntcir/YamamotoLZDZMKO16}. In response to a given ambiguous query, systems in the subtopic discovery task (``intent mining'' or ``query understanding'' in the language of the task organizers) would return a ranked list of subtopics. A subtopic was a query string that disambiguates the search intent of the original query. Assessors manually clustered submitted subtopics into common intents, then voted to assign probabilities to each intent. This process produced a set of weighted subtopics derived from pooled submissions, rather than a fixed set as used in TREC. The main metric, used for both document ranking and subtopic mining, was D$\sharp$-nDCG, a weighted sum of intent recall (I-rec) and a diversified nDCG variant that scales gain values using the subtopic probabilities (D-nDCG)~\cite{sakai-song-dsharp}. The NTCIR tasks made at least three significant strides beyond the TREC web diversity tasks: a method for estimating the probability of a subtopic, metrics that incorporated those probabilities, and an unbounded subtopic set.

The common thread between TREC web and NTCIR IMINE is the focus on covering the possible meanings behind a short web query. While this is certainly one form of coverage, the resource we have built is more strongly focused on deep information needs with multiple aspects.


Recently, TREC Tracks such as NeuCLIR~\cite{neuclir_2024}, RAG~\cite{pradeep2024initialnuggetevaluationresults}, and RAGTIME~\cite{ragtime} have begun revisiting the idea of information coverage as a means of evaluating RAG output, drawing on work from the summarization community.
The Document Understanding Conference (DUC) summarization tasks were the first shared evaluation tasks for text summarization. Systems were provided with a set of newswire documents and produced a summary with a bounded length. The evaluation concentrated on linguistic quality and content coverage~\cite{OVER20071506}. Coverage was measured first by alignment with a manually-written model summary, and then later using the pyramid method~\cite{nenkova-passonneau-2004-evaluating}. In the pyramid method, assessors identified ``summary content units'' (SCUs) that in later years were referred to as ``nuggets''. A summary content unit was an atomic piece of information that an ideal summary would be expected to contain. Multiple assessors created nuggets, and nuggets received a weight corresponding to the number of assessors identifying that nugget, producing the pyramid. The pyramid method was later used in question answering evaluations as well~\cite{lin-demner-fushman-2006-will}, and was the direct antecedent for nugget-based evaluations used in the TREC NeuCLIR track report pilot, RAG, and RAGTIME tracks.





When one has deep information needs, it is intuitive to focus on factual aspects of an information need; however, other types of aspects may be of interest.
The 2022 TREC Fair Ranking track \cite{fair_ranking_2022} evaluates IR systems on fairness with respect to protected or sensitive attributes. Systems retrieve and rank documents while ensuring equitable exposure and representation across different groups, such as subject, geography, or gender. 
%
%
Another example is the 2020 TREC Conversational Assistance Track (CAsT) \cite{cast2020}, which evaluates conversational IR systems that can handle multi-turn, context-dependent queries. In this track, users engage in a sequence of related questions, and systems must interpret each query in the context of the conversation and retrieve relevant passages from large document collections. Each related question can be thought of as an aspect of information coverage.

In this work, we incorporate different types of coverage definitions to form CoverageBench. In the next section, we discuss our construction process in detail. 

\section{CoverageBench} \label{construct}

\begin{table*}[t]
\caption{Dataset Statistics. The average number of words in each document was tokenized by white space. For NeuCLIR and RAGTIME, we report the counts on machine-translated documents, which are all in English for easy comparison. }\label{tab:dataset-stats}
\centering
\begin{tabular}{lrrrrrrr}
\toprule
& \multicolumn{2}{c}{\textbf{Documents / Passages}} & \textbf{Queries} & \multicolumn{4}{c}{\textbf{Nuggets per Query}} \\
\cmidrule(lr){2-3} \cmidrule(lr){4-4} \cmidrule(lr){5-8}
\textbf{Dataset} & \textbf{Count} & \textbf{Avg Words}  & \textbf{Count} & \textbf{Avg} & \textbf{Med} & \textbf{Min} & \textbf{Max} \\
\midrule
CAsT   2020         &  38,429,852 & 59.6  & 25  &  6.1 &    6 &  3 & 12 \\
Fair Ranking 2022   &   6,475,537 & 479.2 & 50  & 29.7 &   26 &  3 & 62 \\
NeuCLIR 2024 Pilot  &  10,038,768 & 348.0 & 19  & 14.9 &   15 & 10 & 25 \\
RAG 2024            & 113,520,750 & 166.7 & 56  & 13.9 &   14 &  6 & 20 \\
RAGTIME 2025        &   4,000,380 & 404.0 & 34  & 15.6 & 16.5 &  8 & 20 \\
CRUX-MultiNews      & \multirow{2}{*}{565,015} & \multirow{2}{*}{88.9}  
                                          & 100 & 14.2 & 14 & 12 & 15 \\
CRUX-DUC04          &             &       & 50  &  7.8 &    8 &  1 & 10 \\
\bottomrule
\end{tabular}
\end{table*}


CoverageBench consists of the resources needed to evaluate information coverage and other artifacts to simplify experimentation (e.g., evaluation code and pre-built indexes). It contains the following components: 


\begin{itemize}[leftmargin=*]
    \item \textit{Topics}. A set of topics (queries)\footnote{We use topic and query interchangeably in this paper.} was drawn from the original task. For datasets where the original topics are complex or multi-faceted (such as conversational turns in CAsT), or lacked context (list of keywords in Fair Ranking) we apply modest modifications (see Section \ref{augment}). The benchmark contains 334 topics. 
    \item \textit{Relevance Labels}. We include relevance judgments (qrels) for each collection. Some datasets required additional LLM judgments. These labels enable relevance and coverage evaluation.
    \item \textit{Document Collection}. All documents from the datasets used in this benchmark are publicly available. We provide instructions on HuggingFace on how to access all document collections.
    \item \textit{Nuggets and Subtopics}. The core augmentation in CoverageBench is a set of nuggets (or subtopics) for each topic, representing the discrete information units that a comprehensive response should cover. For datasets that already include nugget annotations (NeuCLIR, RAG), we adopt or adapt the existing nuggets. For datasets without them (Fair Ranking, CAsT), we derive nuggets through the augmentation process described in Section~\ref{augment}. Each nugget is associated with the set of documents in the collection that contain the corresponding information, enabling evaluation of both 
    retrieval and generation coverage (which nuggets appear in the retrieved set and generated responses, respectively)
    \item \textit{Baselines}. We release baseline ranked lists for each dataset at two stages of the retrieval pipeline. First, initial retrieval of the top-100 results from BM25 \cite{pyserini} and Qwen3-8B \cite{qwen3embedding}. Then, results from reranking the top-50 of the initial retrieval with Rank1-7B \cite{rank1} and Qwen3-Reranker-8B \cite{qwen3embedding}. We release 6 baselines for each dataset, 2 initial retrieval with 4 reranked runs.
\end{itemize}

\subsection{Source Datasets and Augmentation}\label{augment}

We describe each of the datasets in CoverageBench. 
While some datasets contained all the necessary aspects, others required
dataset augmentation to provide each component for each dataset. 
A summary of the dataset statistics is presented in \autoref{tab:dataset-stats}.

\subsubsection{NeuCLIR}

The 2024 TREC NeuCLIR Report Generation pilot task \cite{neuclir_2024} evaluates systems’ ability to generate English reports grounded in multilingual evidence. Systems receive an English report request and must produce a coherent report whose factual claims are supported by citations to relevant news documents written in Chinese, Persian, or Russian. 
For completeness, we include the 19 topics that were judged in all three languages. 
While this dataset is already organized in a manner compatible with coverage evaluation, the number of topics is too small on its own for robust benchmarking.

\subsubsection{RAG}
In the 2024 TREC RAG task~\cite{pradeep2024ragnarokreusableragframework, pradeep2024initialnuggetevaluationresults, thakur2025supportevaluationtrec2024} participants receive non-factoid topic descriptions and the MS MARCO v2.1 segmented collections \cite{pradeep2024ragnarokreusableragframework}, and must retrieve relevant segments and generate structured answers with explicit citations to supporting evidence. 

The original dataset provides nugget annotations for each topic, but the nugget-to-document alignment was never fully recorded or assessed during the original evaluation. We derive nugget-level qrels by using Llama-3.3-70B-Instruct \cite{llama3} as an LLM judge to assess each relevant document against each nugget in the topic. For each topic, nugget, and document triple, the judge determines whether the document contains information corresponding to the nugget. When treating a document as relevant if it contains a nugget, our LLM judge achieves a precision of 69\% and a recall of 90\% across all 56 topics with respect to the original RAG24 relevance judgments. 
We consider this level of agreement reasonable for the purposes of coverage evaluation.\footnote{Since RAG24 did not publish an overview paper, we directly communicated with TREC organizers to verify our approach. We will release the prompt and judged labels upon publication.}

\subsubsection{RAGTIME}
The 2025 TREC RAGTIME Track~\cite{ragtime} hosts a long-form cross-language report generation task, where systems are expected to retrieve information from a multilingual collection consisting of documents in Chinese, Russian, Arabic, and English based on a rich description of information need, i.e., the problem statement and user background, and synthesize a report of several paragraphs long as the response.
The 2025 RAGTIME dataset comes with human-curated nuggets in the form of question and answer pairs, which were extracted from pooled documents created on the track submissions, identical to the 2024 NeuCLIR Report Generation Pilot. 
However, unlike NeuCLIR, there is no explicit annotation on which documents support or contain those nuggets, preventing us from directly applying this dataset for evaluating retrieval coverage. Such information can only be inferred from the sentence support (i.e., whether the report sentence is supported by its citation) and the nugget alignment (i.e., which nugget the report sentence correctly covers) annotations of the evaluated submissions. 

To fill the gap, we use Llama-3.3-70B-Instruct \cite{llama3} with a prompt taking in the problem statement, the nugget (both question and answer), and the document content as the input and ask the model \textit{Does the document contain the specified answer of the question? (Yes/No)} with conditional decoding on the token \texttt{Yes} and \texttt{No} to get an explicit binary answer. 
To ensure completeness, we pooled the top 20 documents from each retrieval run submission in RAGTIME, as well as the top 2 citations from each sentence in the report generation submission, resulting in 22,344 documents in all topic pools. 
We judged all documents in each topic pool against all nuggets, resulting in 686,576 judgments, of which 38,108 came back positive. 

We validate these predictions against the incomplete document-to-nugget annotation extracted from the official annotation on the submissions. 
Specifically, we take all sentences in the report generation submission that are annotated as fully supported by the citations, with the nuggets that they cover, and link the cited documents with the nuggets. Since a document that supports a sentence that covers a specific nugget must also contain that nugget, these inferred support document sets are not complete. However, investigating recall is still meaningful here.
Our LLM-generated labels achieve 72\% recall on this inferred support set. 
We further drew a sample of 60 document-nugget pairs from the false-negative set (i.e., TREC assessors said support but LLM said no) to inspect the quality. 
Of which, we found only 17 pairs (28\%) that the document really supports the nugget, indicating that the true recall should be higher. 
We use these LLM-generated labels to create the nugget-based qrels for evaluation. 

\subsubsection{Fair Ranking}
The 2022 TREC Fair Ranking track \cite{fair_ranking_2022} evaluates IR systems on fairness 
based on the demographic attributes annotated in the collection
While its original structure was not well-suited for measuring coverage metrics such as $\alpha$-nDCG, we adapt the TREC Fair Ranking Track by making use of the annotated corpus where each document has demographic attributes (such as subject geography, popularity). 
In the original task, systems were evaluated on whether their rankings distribute attention across these attribute groups according to a target distribution, measured by AWRF. The original queries are short keyword lists intended for ad hoc retrieval.
We adapt this dataset for coverage evaluation through three transformations: query rewriting, nugget derivation, and relevance label construction.

\begin{itemize}[leftmargin=*]
\item \textit{Queries.} We rewrite each keyword list query into a natural language information-seeking query that explicitly targets coverage across the demographic facets annotated in the dataset. For example, the keyword query for a topic on architecture is rewritten as:
\textit{Overview of architecture and architectural styles from diverse world regions across different time periods including both famous and lesser-known architects.}
The system must now retrieve documents that collectively span the facets of subject geography and popularity embedded in the query.

\item \textit{Nuggets.} We derive nuggets directly from the demographic attribute annotations in the original dataset. Each unique attribute value becomes a nugget that the retrieved set should cover. For the architecture example above, the nuggets include geographic regions (for example, South America, South-eastern Asia, Southern Africa) drawn from the subject geography facet, and popularity levels (High, Low, Medium-High, Medium-Low) drawn from the popularity facet. 

\item \textit{Relevance Label.} The original dataset provides relevance judgments alongside document-level demographic annotations. We combine these to construct nugget-level qrels. For each rewritten query, we identify the set of target nuggets (the facet values the query seeks to cover) and retain only the relevant documents whose demographic annotations match at least one target nugget. Documents that are relevant to the original topic but do not cover any target nugget are excluded, ensuring that the qrels measure coverage of the intended facets.

\end{itemize}

\begin{figure}[t]
    \centering
    \includegraphics[width=1\linewidth]{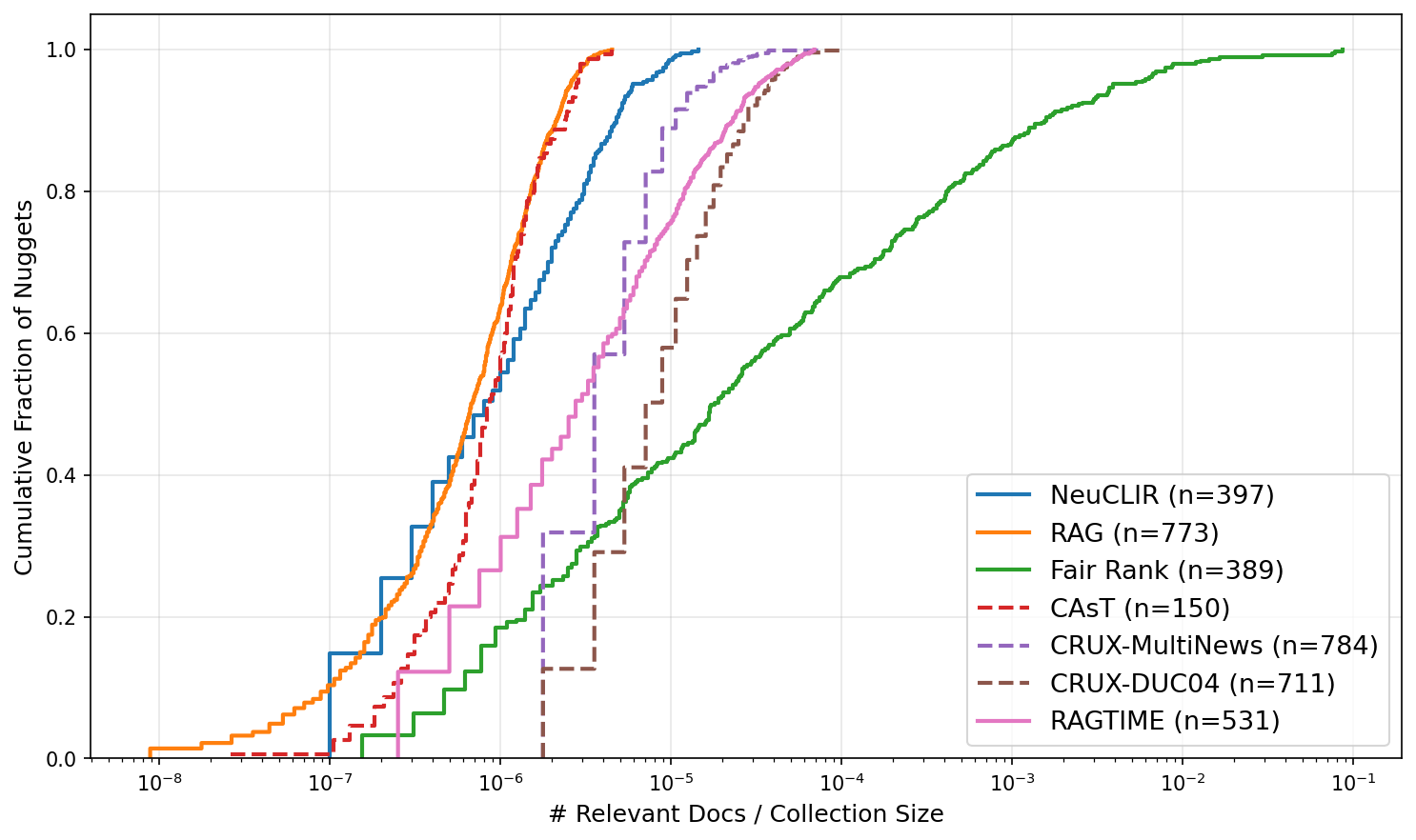}
    \caption{Cumulative distribution of the number of relevant documents per nugget across CoverageBench datasets. Curves further to the left indicate harder datasets where nuggets are covered by fewer documents.}
    \label{fig:cdf}
    \vspace{-1em}
\end{figure}

\subsubsection{CAsT}
The 2020 TREC Conversational Assistance Track (CAsT) \cite{cast2020} was made to evaluate conversational IR systems that can handle multi-turn, context-dependent queries. CAsT emphasizes context modeling, requiring systems to resolve references and retrieve information that satisfies the evolving information need. 
%
Each topic consists of a sequence of related turns, with each turn representing a follow-up question within a broader information need. The original dataset was designed for passage-level conversational retrieval, not coverage evaluation. There are no nugget annotations, and the relevance judgments are sparse, covering only a small number of assessed documents per turn.
We adapt CAsT for coverage evaluation by the following steps:

\begin{itemize}[leftmargin=*]
    \item \textit{Turn Curation.} Not all conversational turns contribute meaningfully to the overarching information need of a topic. Some turns are redundant, revisiting aspects already addressed by earlier turns, while others drift into tangential subtopics. We manually curate the turn sequences by removing redundant and irrelevant turns, retaining only those that represent distinct facets of the topic's information need. 
    \item \textit{Nuggets.} We treat each retained turn as a subtopic of the overarching topic query. The turn's manually rewritten utterance~\cite{cast2020} serves as the subtopic description, and its canonical result document provides a reference answer.
    \item \textit{Relevance Augmentation.} The original CAsT relevance judgments assess only a small pool of documents per turn, which is insufficient for reliable coverage evaluation. We augment the relevance judgments using an LLM judge: for each turn, we use a Llama-3.3-70B-Instruct~\cite{llama3} to assess whether candidate documents from an expanded pool contain information relevant to that turn's subtopic. Documents judged as relevant are added to the nugget-level qrels. Of the 153 canonical relevant documents from the CAsT 2020 ground truth, 86 (56.2\%) were recovered in the expanded judgment pool. Examining the missed documents by hand, we found that almost all were not relevant, so the true recall should be higher.
\end{itemize}




\subsubsection{CRUX-MultiNews and CRUX-DUC04}

Controlled Retrieval-augmented Context Evaluation (CRUX)~\cite{crux} is an evaluation framework designed to assess retrieval contexts in long-form RAG scenarios, which does not require augmentation for CoverageBench. 
Rather than evaluating retrieval purely through relevance-based ranking metrics, CRUX measures how completely a retrieval context covers the information needed for long-form generation, using question-based evaluation. 

The framework is built on multi-document summarization datasets, Multi-News~\cite{multinews} and DUC-2004~\cite{DUC04}, where human-written summaries serve as oracle long-form results that define the scope of relevant retrieval context. From each summary, CRUX generates an open-ended query, a diverse set of knowledge-intensive sub-questions, and decontextualized passage-level chunks from the source documents. An LLM judges whether each passage answers each sub-question, producing a matrix of graded answerability scores that supports fine-grained coverage and density metrics. Because these datasets already provide mappings between nuggets and supporting passages, they are compatible with coverage evaluation and require no further adaptation for inclusion in our benchmark.








\subsection{Dataset Statistics}

Table \ref{tab:dataset-stats} summarizes the key statistics of the datasets in CoverageBench. The benchmark contains 334 topics in total. The document collection spans from 565k passages (CRUX-MultiNews and CRUX-DUC04) to over 113M segments (RAG). The datasets vary in the number of queries, from 19 (NeuCLIR) to 100 (CRUX-MultiNews). 

The number of nuggets per query also varies within and across datasets. Fair Ranking has the highest average nugget count (29.7/query). and the widest range (3 to 62). CAsT has the fewest nuggets per query on average (6.1). This is expected as the nuggets are based on a conversational structure, and each nugget is equivalent to a turn in the conversation. 
Document length also shapes part of the coverage challenge. A longer document is likely to cover more nuggets. CAsT passages were the shortest, averaging 59.6 words per passage. Fair Ranking and RAGTIME documents are longer, 479 and 404 words on average, respectively. 


Figure \ref{fig:cdf} shows the cumulative distribution of the number of relevant documents per nugget, normalized by the collection size, for all of the benchmark datasets. Curves to the left indicate datasets where each nugget is supported by a smaller fraction of the collection. Fair Ranking has the most queries per nugget, so individual nuggets are well-attested in a collection. Here, the challenge is in covering all of the nuggets in a ranked list since it has the highest average of nuggets per query. 

\begin{table*}[t]
\centering
\caption{Baseline retrieval and reranking relevance effectiveness measured by nDCG@20. The average column reports the macro-average over all seven datasets. }
\label{tab:baseline-ndcg-20}
\vspace{-1em}
\begin{tabular}{ll cccccccc}
\toprule
{Initial} & {Rerank} &{NeuCLIR} & {RAG} & {Fair Rank} & {CAsT} & {CRUX-MultiNews} & {CRUX-DUC04}& {RAGTIME} & {Avg} \\
\midrule
BM25      & --      & 0.328 & 0.599 & 0.143 & 0.475 & 0.429 & 0.448 & 0.596 & 0.431 \\
Qwen3-8B  & --      & 0.819 & 0.854 & 0.094 & {0.700} & 0.610 & 0.701 & 0.774 & 0.650 \\
\midrule
BM25      & Rank1   & 0.520 & 0.758 & \textbf{0.161} & 0.584 & 0.533 & 0.610 & 0.624 & 0.541 \\
Qwen3-8B  & Rank1   & 0.821 & \textbf{0.915} & 0.124 & \textbf{0.704} & 0.616 & \textbf{0.754} & \textbf{0.800} & \textbf{0.676} \\
BM25      & Qwen3-R & 0.581 & 0.777 & 0.103 & 0.420 & 0.494 & 0.524 & 0.730 & 0.518 \\
Qwen3-8B  & Qwen3-R & \textbf{0.860} & 0.898 & 0.052 & 0.675 & \textbf{0.620} & 0.713 & 0.609 & 0.632 \\
\bottomrule
\end{tabular}
\end{table*}

\begin{table*}[t]
\centering
\caption{Baseline retrieval and reranking results on coverage: $\alpha$-nDCG@20 and Subtopic Recall (StRecall@20)}
\label{tab:baseline-diversity-20}
\vspace{-1em}
\setlength{\tabcolsep}{2.5pt}
\begin{tabular}{ll cccccccc cccccccc}
\toprule
& & \multicolumn{8}{c}{{$\alpha$-nDCG@20}} & \multicolumn{8}{c}{{StRecall@20}} \\
\cmidrule(lr){3-10} \cmidrule(lr){11-18}
{Initial} & {Rerank}
& Neu & RAG & FR & CAsT & CX-M & CX-D & RAGT & Avg & Neu & RAG & FR & CAsT & CX-M & CX-D & RAGT & Avg \\
\midrule
BM25      & --      & 0.349 & 0.450 & 0.109 & 0.357 & 0.469 & 0.476 & 0.486 & 0.385 & 0.545 & 0.708 & 0.171 & 0.577 & 0.634 & 0.659 & 0.664 & 0.565 \\
Qwen3-8B  & --      & 0.627 & 0.683 & 0.074 & 0.437 & 0.648 & 0.652 & 0.531 & 0.522 & 0.836 & 0.899 & 0.117 & 0.597 & 0.813 & 0.814 & 0.707 & 0.683 \\
\midrule
BM25      & Rank1   & 0.538 & 0.658 & \textbf{0.122} & 0.429 & 0.583 & 0.621 & 0.527 & 0.497 & 0.664 & 0.818 & \textbf{0.217} & \textbf{0.639} & 0.710 & 0.732 & 0.690 & 0.639 \\
Qwen3-8B  & Rank1   & 0.613 & \textbf{0.742} & 0.090 & \textbf{0.440} & 0.634 & \textbf{0.690} & \textbf{0.590} & \textbf{0.543} & 0.786 & \textbf{0.935} & 0.133 & 0.619 & \textbf{0.831} & \textbf{0.839} & \textbf{0.773} & \textbf{0.702} \\
BM25      & Qwen3-R & 0.583 & 0.633 & 0.065 & 0.321 & 0.551 & 0.554 & 0.518 & 0.461 & 0.684 & 0.819 & 0.143 & 0.576 & 0.689 & 0.696 & 0.715 & 0.617 \\
Qwen3-8B  & Qwen3-R & \textbf{0.691} & 0.705 & 0.036 & 0.407 & \textbf{0.652} & 0.663 & 0.490 & 0.521 & \textbf{0.839} & 0.903 & 0.060 & 0.600 & 0.817 & 0.826 & 0.660 & 0.672 \\
\bottomrule
\end{tabular}
\end{table*}

\section{Baseline Experiments}

To characterize the coverage properties of each dataset and provide reference points for future work, we evaluate a set of retrieval and reranking configurations across all seven CoverageBench datasets.

\subsection{Setup}
We evaluate two initial retrieval models representing sparse and dense retrieval paradigms. BM25 \cite{pyserini} serves as the sparse baseline, using default parameters. Qwen3-8B \cite{qwen3embedding} serves as the dense baseline, encoding queries and documents into dense vectors for nearest-neighbor retrieval with FAISS~\cite{faiss}. 
%
%
We apply two rerankers to each initial retrieval run to rerank the top 50: Rank1-7B \cite{rank1} and Qwen3-Reranker-8B \cite{qwen3embedding}. For NeuCLIR and RAGTIME, we rerank with the machine-translated English documents since the datasets are multilingual.  Combined with the two initial runs, we obtain a total of six retrieval configurations per dataset. 

All results are evaluated using nDCG (for relevance), $\alpha$-nDCG, and Subtopic Recall (StRecall) with a rank cutoff of 20.


\begin{figure}
    \centering
    \includegraphics[width=1\linewidth]{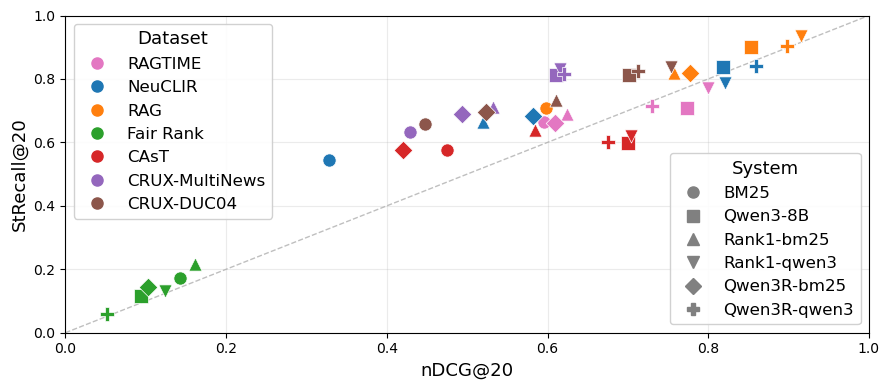}
    \vspace{-2.5em}
    \caption{nDCG@20 vs.\ Subtopic Recall@20}
    \label{fig:scatterplot}
    \vspace{-1em}
\end{figure}

\subsection{Results}


In Tables~\ref{tab:baseline-ndcg-20} and \ref{tab:baseline-diversity-20}, across most datasets, configurations with Qwen3-8B as the initial rank performed better than BM25 in both relevance and coverage, which aligns with findings in other works in dense retrieval. The exception is with Fair Ranking, where BM25 performs better across all three metrics. This is likely because the rewritten queries contain explicit facet keywords, which would favor lexical matches. The original retrieval task in Fair Ranking is also more syntactic than semantic, which favors surface form matching. 

On NeuCLIR, Rank1 applied to Qwen3-8B matches the initial run on nDCG (0.821 vs.\ 0.819) but drops StRecall from 0.836 to 0.786, indicating that the reranking starts to exploit particular aspects of relevance for optimizing relevance but misses other aspects in the top 20 documents. 

The highest scoring system on relevance does not always achieve the highest StRecall. On CRUX-MultiNews, Qwen3-Reranker on Qwen3-8B leads on nDCG@20 (0.620) while Rank1 on Qwen3-8B leads on StRecall@20 (0.831). 
This demonstrates how CoverageBench provides a different perspective than evaluating only on relevance-based metrics. 
CAsT presents a coverage ceiling that no configuration breaks through. As Figure~\ref{fig:scatterplot} shows, all six configurations cluster around a StRecall of approximately 0.55, despite spanning nDCG scores from 0.323 to 0.651. 
\section{Conclusion}

We introduced CoverageBench, a benchmark for evaluating information coverage in retrieval systems. Our work demonstrates that coverage evaluation can be derived from existing retrieval collections without the cost of creating entirely new test collections from scratch. Baseline experiments across six retrieval configurations reveal variation in coverage performance both across datasets and methods. These results demonstrate the need for continued research in coverage-aware retrieval, particularly as RAG systems increasingly depend on comprehensive information gathering rather than simple relevance matching. All topics, nuggets, relevance labels, and baseline rankings are publicly available on HuggingFace, enabling the community to advance coverage evaluation research and develop retrieval systems better suited for applications where comprehensive information access matters.

\subsection*{Disclaimer}
Certain products are named in this paper in order to fully specify the experimental procedure adequately. Such mentions should not be taken as endorsement or recommendation of any company, product, or service by NIST, nor are they intended to imply that the products identified are necessarily the best available for this purpose.


\balance
\bibliographystyle{ACM-Reference-Format}
\bibliography{citations}

@article{faiss,
      title={The Faiss library},
      author={Matthijs Douze and Alexandr Guzhva and Chengqi Deng and Jeff Johnson and Gergely Szilvasy and Pierre-Emmanuel Mazaré and Maria Lomeli and Lucas Hosseini and Hervé Jégou},
      year={2024},
      eprint={2401.08281},
      archivePrefix={arXiv},
      primaryClass={cs.LG}
}

@misc{ragtime,
      title={Overview of the TREC 2025 RAGTIME Track}, 
      author={Dawn Lawrie and Sean MacAvaney and James Mayfield and Luca Soldaini and Eugene Yang and Andrew Yates},
      year={2026},
      eprint={2602.10024},
      archivePrefix={arXiv},
      primaryClass={cs.IR},
      url={https://arxiv.org/abs/2602.10024}, 
}

@inproceedings{DBLP:conf/trec/ClarkeCS09,
  author       = {Charles L. A. Clarke and
                  Nick Craswell and
                  Ian Soboroff},
  editor       = {Ellen M. Voorhees and
                  Lori P. Buckland},
  title        = {Overview of the {TREC} 2009 Web Track},
  booktitle    = {Proceedings of The Eighteenth Text REtrieval Conference, {TREC} 2009,
                  Gaithersburg, Maryland, USA, November 17-20, 2009},
  series       = {{NIST} Special Publication},
  volume       = {500-278},
  publisher    = {National Institute of Standards and Technology {(NIST)}},
  year         = {2009},
  url          = {http://trec.nist.gov/pubs/trec18/papers/WEB09.OVERVIEW.pdf},
  bibsource    = {dblp computer science bibliography, https://dblp.org}
}

@inproceedings{DBLP:conf/trec/ClarkeCSC10,
  author       = {Charles L. A. Clarke and
                  Nick Craswell and
                  Ian Soboroff and
                  Gordon V. Cormack},
  editor       = {Ellen M. Voorhees and
                  Lori P. Buckland},
  title        = {Overview of the {TREC} 2010 Web Track},
  booktitle    = {Proceedings of The Nineteenth Text REtrieval Conference, {TREC} 2010,
                  Gaithersburg, Maryland, USA, November 16-19, 2010},
  series       = {{NIST} Special Publication},
  volume       = {500-294},
  publisher    = {National Institute of Standards and Technology {(NIST)}},
  year         = {2010},
  url          = {https://trec.nist.gov/pubs/trec19/papers/WEB.OVERVIEW.pdf},
  bibsource    = {dblp computer science bibliography, https://dblp.org}
}

@inproceedings{DBLP:conf/trec/ClarkeCSV11,
  author       = {Charles L. A. Clarke and
                  Nick Craswell and
                  Ian Soboroff and
                  Ellen M. Voorhees},
  editor       = {Ellen M. Voorhees and
                  Lori P. Buckland},
  title        = {Overview of the {TREC} 2011 Web Track},
  booktitle    = {Proceedings of The Twentieth Text REtrieval Conference, {TREC} 2011,
                  Gaithersburg, Maryland, USA, November 15-18, 2011},
  series       = {{NIST} Special Publication},
  volume       = {500-296},
  publisher    = {National Institute of Standards and Technology {(NIST)}},
  year         = {2011},
  url          = {http://trec.nist.gov/pubs/trec20/papers/WEB.OVERVIEW.pdf},
  bibsource    = {dblp computer science bibliography, https://dblp.org}
}

@inproceedings{DBLP:conf/wsdm/ClarkeCSA11,
  author       = {Charles L. A. Clarke and
                  Nick Craswell and
                  Ian Soboroff and
                  Azin Ashkan},
  editor       = {Irwin King and
                  Wolfgang Nejdl and
                  Hang Li},
  title        = {A comparative analysis of cascade measures for novelty and diversity},
  booktitle    = {Proceedings of the Forth International Conference on Web Search and
                  Web Data Mining, {WSDM} 2011, Hong Kong, China, February 9-12, 2011},
  pages        = {75--84},
  publisher    = {{ACM}},
  year         = {2011},
  url          = {https://doi.org/10.1145/1935826.1935847},
  doi          = {10.1145/1935826.1935847},
  bibsource    = {dblp computer science bibliography, https://dblp.org}
}

@article{ndcg,
author = {J\"{a}rvelin, Kalervo and Kek\"{a}l\"{a}inen, Jaana},
title = {Cumulated gain-based evaluation of IR techniques},
year = {2002},
issue_date = {October 2002},
publisher = {Association for Computing Machinery},
address = {New York, NY, USA},
volume = {20},
number = {4},
issn = {1046-8188},
url = {https://doi.org/10.1145/582415.582418},
doi = {10.1145/582415.582418},
journal = {ACM Trans. Inf. Syst.},
month = oct,
pages = {422–446},
numpages = {25}
}

@inproceedings{qa_track,
    title = "The {TREC}-8 Question Answering Track",
    author = "Voorhees, Ellen M.  and
      Tice, Dawn M.",
    editor = "Gavrilidou, M.  and
      Carayannis, G.  and
      Markantonatou, S.  and
      Piperidis, S.  and
      Stainhauer, G.",
    booktitle = "Proceedings of the Second International Conference on Language Resources and Evaluation ({LREC}{'}00)",
    month = may,
    year = "2000",
    address = "Athens, Greece",
    publisher = "European Language Resources Association (ELRA)",
    url = "https://aclanthology.org/L00-1018/"
}

@book{intro_to_ir,
  author    = {Christopher D. Manning and Prabhakar Raghavan and Hinrich Schütze},
  title     = {Introduction to Information Retrieval},
  publisher = {Cambridge University Press},
  year      = {2008}
}

@inproceedings{DBLP:conf/ntcir/Song0SKLSWO11,
  author       = {Ruihua Song and
                  Min Zhang and
                  Tetsuya Sakai and
                  Makoto P. Kato and
                  Yiqun Liu and
                  Miho Sugimoto and
                  Qinglei Wang and
                  Naoki Orii},
  editor       = {Noriko Kando and
                  Daisuke Ishikawa and
                  Miho Sugimoto},
  title        = {Overview of the {NTCIR-9} {INTENT} Task},
  booktitle    = {Proceedings of the 9th {NTCIR} Workshop Meeting on Evaluation of Information
                  Access Technologies: Information Retrieval, Question Answering and
                  Cross-Lingual Information Access, NTCIR-9, National Center of Sciences,
                  Tokyo, Japan, December 6-9, 2011},
  publisher    = {National Institute of Informatics {(NII)}},
  year         = {2011},
  url          = {http://research.nii.ac.jp/ntcir/workshop/OnlineProceedings9/NTCIR/01-NTCIR9-OV-INTENT-SongR.pdf},
  bibsource    = {dblp computer science bibliography, https://dblp.org}
}

@inproceedings{DBLP:conf/sigir/SakaiDYLZKSI13,
  author       = {Tetsuya Sakai and
                  Zhicheng Dou and
                  Takehiro Yamamoto and
                  Yiqun Liu and
                  Min Zhang and
                  Makoto P. Kato and
                  Ruihua Song and
                  Mayu Iwata},
  editor       = {Gareth J. F. Jones and
                  Paraic Sheridan and
                  Diane Kelly and
                  Maarten de Rijke and
                  Tetsuya Sakai},
  title        = {Summary of the {NTCIR-10} {INTENT-2} task: subtopic mining and search
                  result diversification},
  booktitle    = {The 36th International {ACM} {SIGIR} conference on research and development
                  in Information Retrieval, {SIGIR} '13, Dublin, Ireland - July 28 -
                  August 01, 2013},
  pages        = {761--764},
  publisher    = {{ACM}},
  year         = {2013},
  url          = {https://doi.org/10.1145/2484028.2484104},
  doi          = {10.1145/2484028.2484104},
  bibsource    = {dblp computer science bibliography, https://dblp.org}
}

@inproceedings{DBLP:conf/ntcir/LiuS0DYKOZ14,
  author       = {Yiqun Liu and
                  Ruihua Song and
                  Min Zhang and
                  Zhicheng Dou and
                  Takehiro Yamamoto and
                  Makoto P. Kato and
                  Hiroaki Ohshima and
                  Ke Zhou},
  editor       = {Noriko Kando and
                  Hideo Joho and
                  Kazuaki Kishida},
  title        = {Overview of the {NTCIR-11} IMine Task},
  booktitle    = {Proceedings of the 11th {NTCIR} Conference on Evaluation of Information
                  Access Technologies, NTCIR-11, National Center of Sciences, Tokyo,
                  Japan, December 9-12, 2014},
  publisher    = {National Institute of Informatics {(NII)}},
  year         = {2014},
  url          = {http://research.nii.ac.jp/ntcir/workshop/OnlineProceedings11/pdf/NTCIR/OVERVIEW/01-NTCIR11-OV-IMINE-LiuY.pdf},
  bibsource    = {dblp computer science bibliography, https://dblp.org}
}

@inproceedings{DBLP:conf/ntcir/YamamotoLZDZMKO16,
  author       = {Takehiro Yamamoto and
                  Yiqun Liu and
                  Min Zhang and
                  Zhicheng Dou and
                  Ke Zhou and
                  Ilya Markov and
                  Makoto P. Kato and
                  Hiroaki Ohshima and
                  Sumio Fujita},
  editor       = {Noriko Kando and
                  Tetsuya Sakai and
                  Mark Sanderson},
  title        = {Overview of the {NTCIR-12} IMine-2 Task},
  booktitle    = {Proceedings of the 12th {NTCIR} Conference on Evaluation of Information
                  Access Technologies, National Center of Sciences, Tokyo, Japan, June
                  7-10, 2016},
  publisher    = {National Institute of Informatics {(NII)}},
  year         = {2016},
  url          = {http://research.nii.ac.jp/ntcir/workshop/OnlineProceedings12/pdf/ntcir/OVERVIEW/01-NTCIR12-OV-IMINE-YamamotoT.pdf},
  bibsource    = {dblp computer science bibliography, https://dblp.org}
}

@article{bm25_and_beyond_statistical,
author = {Robertson, Stephen and Zaragoza, Hugo},
title = {The Probabilistic Relevance Framework: BM25 and Beyond},
year = {2009},
issue_date = {April 2009},
publisher = {Now Publishers Inc.},
address = {Hanover, MA, USA},
volume = {3},
number = {4},
issn = {1554-0669},
url = {https://doi.org/10.1561/1500000019},
doi = {10.1561/1500000019},
journal = {Found. Trends Inf. Retr.},
month = apr,
pages = {333–389},
numpages = {57}
}

@misc{colbert,
      title={ColBERT: Efficient and Effective Passage Search via Contextualized Late Interaction over BERT}, 
      author={Omar Khattab and Matei Zaharia},
      year={2020},
      eprint={2004.12832},
      archivePrefix={arXiv},
      primaryClass={cs.IR},
      url={https://arxiv.org/abs/2004.12832}, 
}

@misc{spladev2,
      title={SPLADE v2: Sparse Lexical and Expansion Model for Information Retrieval}, 
      author={Thibault Formal and Carlos Lassance and Benjamin Piwowarski and Stéphane Clinchant},
      year={2021},
      eprint={2109.10086},
      archivePrefix={arXiv},
      primaryClass={cs.IR},
      url={https://arxiv.org/abs/2109.10086}, 
}

@inproceedings{novelty_diversity_IR_eval,
author = {Clarke, Charles L.A. and Kolla, Maheedhar and Cormack, Gordon V. and Vechtomova, Olga and Ashkan, Azin and B\"{u}ttcher, Stefan and MacKinnon, Ian},
title = {Novelty and diversity in information retrieval evaluation},
year = {2008},
isbn = {9781605581644},
publisher = {Association for Computing Machinery},
address = {New York, NY, USA},
url = {https://doi.org/10.1145/1390334.1390446},
doi = {10.1145/1390334.1390446},
booktitle = {Proceedings of the 31st Annual International ACM SIGIR Conference on Research and Development in Information Retrieval},
pages = {659–666},
numpages = {8},
location = {Singapore, Singapore},
series = {SIGIR '08}
}

@inproceedings{beyond_independent_relevance,
author = {Zhai, Cheng Xiang and Cohen, William W. and Lafferty, John},
title = {Beyond independent relevance: methods and evaluation metrics for subtopic retrieval},
year = {2003},
isbn = {1581136463},
publisher = {Association for Computing Machinery},
address = {New York, NY, USA},
url = {https://doi.org/10.1145/860435.860440},
doi = {10.1145/860435.860440},
booktitle = {Proceedings of the 26th Annual International ACM SIGIR Conference on Research and Development in Informaion Retrieval},
pages = {10–17},
numpages = {8},
location = {Toronto, Canada},
series = {SIGIR '03}
}

@inproceedings{use_of_MMR,
author = {Carbonell, Jaime and Goldstein, Jade},
title = {The use of MMR, diversity-based reranking for reordering documents and producing summaries},
year = {1998},
isbn = {1581130155},
publisher = {Association for Computing Machinery},
address = {New York, NY, USA},
url = {https://doi.org/10.1145/290941.291025},
doi = {10.1145/290941.291025},
booktitle = {Proceedings of the 21st Annual International ACM SIGIR Conference on Research and Development in Information Retrieval},
pages = {335–336},
numpages = {2},
location = {Melbourne, Australia},
series = {SIGIR '98}
}

@misc{rag_knowledge_intensive_nlp,
      title={Retrieval-Augmented Generation for Knowledge-Intensive NLP Tasks}, 
      author={Patrick Lewis and Ethan Perez and Aleksandra Piktus and Fabio Petroni and Vladimir Karpukhin and Naman Goyal and Heinrich Küttler and Mike Lewis and Wen-tau Yih and Tim Rocktäschel and Sebastian Riedel and Douwe Kiela},
      year={2021},
      eprint={2005.11401},
      archivePrefix={arXiv},
      primaryClass={cs.CL},
      url={https://arxiv.org/abs/2005.11401}, 
}

@misc{rag_survey,
      title={Retrieval-Augmented Generation for Large Language Models: A Survey}, 
      author={Yunfan Gao and Yun Xiong and Xinyu Gao and Kangxiang Jia and Jinliu Pan and Yuxi Bi and Yi Dai and Jiawei Sun and Meng Wang and Haofen Wang},
      year={2024},
      eprint={2312.10997},
      archivePrefix={arXiv},
      primaryClass={cs.CL},
      url={https://arxiv.org/abs/2312.10997}, 
}

@inproceedings{power_of_noise, series={SIGIR 2024},
   title={The Power of Noise: Redefining Retrieval for RAG Systems},
   url={http://dx.doi.org/10.1145/3626772.3657834},
   DOI={10.1145/3626772.3657834},
   booktitle={Proceedings of the 47th International ACM SIGIR Conference on Research and Development in Information Retrieval},
   publisher={ACM},
   author={Cuconasu, Florin and Trappolini, Giovanni and Siciliano, Federico and Filice, Simone and Campagnano, Cesare and Maarek, Yoelle and Tonellotto, Nicola and Silvestri, Fabrizio},
   year={2024},
   month=jul, pages={719–729},
   collection={SIGIR 2024} }

@misc{auto_argue,
      title={Auto-ARGUE: LLM-Based Report Generation Evaluation}, 
      author={William Walden and Marc Mason and Orion Weller and Laura Dietz and John Conroy and Neil Molino and Hannah Recknor and Bryan Li and Gabrielle Kaili-May Liu and Yu Hou and Dawn Lawrie and James Mayfield and Eugene Yang},
      year={2025},
      eprint={2509.26184},
      archivePrefix={arXiv},
      primaryClass={cs.IR},
      url={https://arxiv.org/abs/2509.26184}, 
}

@inproceedings{deconstructing_nuggets,
author = {Lin, Jimmy and Zhang, Pengyi},
title = {Deconstructing nuggets: the stability and reliability of complex question answering evaluation},
year = {2007},
isbn = {9781595935977},
publisher = {Association for Computing Machinery},
address = {New York, NY, USA},
url = {https://doi.org/10.1145/1277741.1277799},
doi = {10.1145/1277741.1277799},
booktitle = {Proceedings of the 30th Annual International ACM SIGIR Conference on Research and Development in Information Retrieval},
pages = {327–334},
numpages = {8},
location = {Amsterdam, The Netherlands},
series = {SIGIR '07}
}

@inproceedings{eval_summary_answer_2_coins,
    title = "Evaluating Summaries and Answers: Two Sides of the Same Coin?",
    author = "Lin, Jimmy  and
      Demner-Fushman, Dina",
    editor = "Goldstein, Jade  and
      Lavie, Alon  and
      Lin, Chin-Yew  and
      Voss, Clare",
    booktitle = "Proceedings of the {ACL} Workshop on Intrinsic and Extrinsic Evaluation Measures for Machine Translation and/or Summarization",
    month = jun,
    year = "2005",
    address = "Ann Arbor, Michigan",
    publisher = "Association for Computational Linguistics",
    url = "https://aclanthology.org/W05-0906/",
    pages = "41--48"
}

@article{OVER20071506,
title = {DUC in context},
journal = {Information Processing \& Management},
volume = {43},
number = {6},
pages = {1506-1520},
year = {2007},
issn = {0306-4573},
doi = {https://doi.org/10.1016/j.ipm.2007.01.019},
url = {https://www.sciencedirect.com/science/article/pii/S0306457307000404},
author = {Paul Over and Hoa Dang and Donna Harman},
}

@misc{neuclir_2024,
      title={Overview of the TREC 2024 NeuCLIR Track}, 
      author={Dawn Lawrie and Sean MacAvaney and James Mayfield and Paul McNamee and Douglas W. Oard and Luca Soldaini and Eugene Yang},
      year={2025},
      eprint={2509.14355},
      archivePrefix={arXiv},
      primaryClass={cs.IR},
      url={https://arxiv.org/abs/2509.14355}, 
}

@misc{pradeep2024initialnuggetevaluationresults,
      title={Initial Nugget Evaluation Results for the TREC 2024 RAG Track with the AutoNuggetizer Framework}, 
      author={Ronak Pradeep and Nandan Thakur and Shivani Upadhyay and Daniel Campos and Nick Craswell and Jimmy Lin},
      year={2024},
      eprint={2411.09607},
      archivePrefix={arXiv},
      primaryClass={cs.IR},
      url={https://arxiv.org/abs/2411.09607}, 
}

@misc{pradeep2024ragnarokreusableragframework,
      title={Ragnar\"ok: A Reusable RAG Framework and Baselines for TREC 2024 Retrieval-Augmented Generation Track}, 
      author={Ronak Pradeep and Nandan Thakur and Sahel Sharifymoghaddam and Eric Zhang and Ryan Nguyen and Daniel Campos and Nick Craswell and Jimmy Lin},
      year={2024},
      eprint={2406.16828},
      archivePrefix={arXiv},
      primaryClass={cs.IR},
      url={https://arxiv.org/abs/2406.16828}, 
}

@misc{thakur2025supportevaluationtrec2024,
      title={Support Evaluation for the TREC 2024 RAG Track: Comparing Human versus LLM Judges}, 
      author={Nandan Thakur and Ronak Pradeep and Shivani Upadhyay and Daniel Campos and Nick Craswell and Jimmy Lin},
      year={2025},
      eprint={2504.15205},
      archivePrefix={arXiv},
      primaryClass={cs.CL},
      url={https://arxiv.org/abs/2504.15205}, 
}

@misc{fair_ranking_2022,
      title={Overview of the TREC 2022 Fair Ranking Track}, 
      author={Michael D. Ekstrand and Graham McDonald and Amifa Raj and Isaac Johnson},
      year={2023},
      eprint={2302.05558},
      archivePrefix={arXiv},
      primaryClass={cs.IR},
      url={https://arxiv.org/abs/2302.05558}, 
}

@inproceedings{cast2020,
  author    = {Jeffrey Dalton and Chenyan Xiong and Jamie Callan},
  title     = {{CAsT} 2020: The Conversational Assistance Track Overview},
  booktitle = {The Twenty-Ninth Text REtrieval Conference Proceedings (TREC 2020)},
  publisher = {National Institute of Standards and Technology (NIST)},
  series    = {NIST Special Publication},
  volume    = {1266},
  year      = {2021},
  url       = {https://trec.nist.gov/pubs/trec29/papers/OVERVIEW.C.pdf}
}

@misc{crux,
      title={Controlled Retrieval-augmented Context Evaluation for Long-form RAG}, 
      author={Jia-Huei Ju and Suzan Verberne and Maarten de Rijke and Andrew Yates},
      year={2026},
      eprint={2506.20051},
      archivePrefix={arXiv},
      primaryClass={cs.IR},
      url={https://arxiv.org/abs/2506.20051}, 
}

@inproceedings{nenkova-passonneau-2004-evaluating,
    title = "Evaluating Content Selection in Summarization: The Pyramid Method",
    author = "Nenkova, Ani  and
      Passonneau, Rebecca",
    booktitle = "Proceedings of the Human Language Technology Conference of the North {A}merican Chapter of the Association for Computational Linguistics: {HLT}-{NAACL} 2004",
    month = may # " 2 - " # may # " 7",
    year = "2004",
    address = "Boston, Massachusetts, USA",
    publisher = "Association for Computational Linguistics",
    url = "https://aclanthology.org/N04-1019/",
    pages = "145--152"
}

@inproceedings{pyserini,
author = {Lin, Jimmy and Ma, Xueguang and Lin, Sheng-Chieh and Yang, Jheng-Hong and Pradeep, Ronak and Nogueira, Rodrigo},
title = {Pyserini: A Python Toolkit for Reproducible Information Retrieval Research with Sparse and Dense Representations},
year = {2021},
isbn = {9781450380379},
publisher = {Association for Computing Machinery},
address = {New York, NY, USA},
url = {https://doi.org/10.1145/3404835.3463238},
doi = {10.1145/3404835.3463238},
booktitle = {Proceedings of the 44th International ACM SIGIR Conference on Research and Development in Information Retrieval},
pages = {2356–2362},
numpages = {7},
location = {Virtual Event, Canada},
series = {SIGIR '21}
}

@misc{qwen3embedding,
      title={Qwen3 Embedding: Advancing Text Embedding and Reranking Through Foundation Models}, 
      author={Yanzhao Zhang and Mingxin Li and Dingkun Long and Xin Zhang and Huan Lin and Baosong Yang and Pengjun Xie and An Yang and Dayiheng Liu and Junyang Lin and Fei Huang and Jingren Zhou},
      year={2025},
      eprint={2506.05176},
      archivePrefix={arXiv},
      primaryClass={cs.CL},
      url={https://arxiv.org/abs/2506.05176}, 
}

@misc{rank1,
      title={Rank1: Test-Time Compute for Reranking in Information Retrieval}, 
      author={Orion Weller and Kathryn Ricci and Eugene Yang and Andrew Yates and Dawn Lawrie and Benjamin Van Durme},
      year={2025},
      eprint={2502.18418},
      archivePrefix={arXiv},
      primaryClass={cs.IR},
      url={https://arxiv.org/abs/2502.18418}, 
}

@inproceedings{lin-demner-fushman-2006-will,
    title = "Will Pyramids Built of Nuggets Topple Over?",
    author = "Lin, Jimmy  and
      Demner-Fushman, Dina",
    editor = "Moore, Robert C.  and
      Bilmes, Jeff  and
      Chu-Carroll, Jennifer  and
      Sanderson, Mark",
    booktitle = "Proceedings of the Human Language Technology Conference of the {NAACL}, Main Conference",
    month = jun,
    year = "2006",
    address = "New York City, USA",
    publisher = "Association for Computational Linguistics",
    url = "https://aclanthology.org/N06-1049/",
    pages = "383--390"
}

@inproceedings{sakai-song-dsharp,
author = {Sakai, Tetsuya and Song, Ruihua},
title = {Evaluating diversified search results using per-intent graded relevance},
booktitle = {Proceedings of SIGIR 2011},
year = {2011},
isbn = {9781450307574},
publisher = {Association for Computing Machinery},
address = {New York, NY, USA},
url = {https://doi.org/10.1145/2009916.2010055},
doi = {10.1145/2009916.2010055},
pages = {1043–1052},
numpages = {10},
location = {Beijing, China},
series = {SIGIR '11}
}

@inproceedings{DBLP:conf/trec/Over97,
  author       = {Paul Over},
  editor       = {Ellen M. Voorhees and
                  Donna K. Harman},
  title        = {{TREC-6} Interactive Report},
  booktitle    = {Proceedings of The Sixth Text REtrieval Conference, {TREC} 1997, Gaithersburg,
                  Maryland, USA, November 19-21, 1997},
  series       = {{NIST} Special Publication},
  volume       = {500-240},
  pages        = {73--81},
  publisher    = {National Institute of Standards and Technology {(NIST)}},
  year         = {1997},
  url          = {http://trec.nist.gov/pubs/trec6/papers/t6irep.ps},
  bibsource    = {dblp computer science bibliography, https://dblp.org}
}

@article{intent_based_metrics,
author = {Chapelle, Olivier and Ji, Shihao and Liao, Ciya and Velipasaoglu, Emre and Lai, Larry and Wu, Su-Lin},
title = {Intent-based diversification of web search results: metrics and algorithms},
year = {2011},
issue_date = {Dec 2011},
publisher = {Kluwer Academic Publishers},
address = {USA},
volume = {14},
number = {6},
issn = {1386-4564},
journal = {Inf. Retr.},
month = dec,
pages = {572–592},
numpages = {21}
}

@inproceedings{multinews,
    title = "Multi-News: A Large-Scale Multi-Document Summarization Dataset and Abstractive Hierarchical Model",
    author = "Fabbri, Alexander  and
      Li, Irene  and
      She, Tianwei  and
      Li, Suyi  and
      Radev, Dragomir",
    editor = "Korhonen, Anna  and
      Traum, David  and
      M{\`a}rquez, Llu{\'i}s",
    booktitle = "Proceedings of the 57th Annual Meeting of the Association for Computational Linguistics",
    month = jul,
    year = "2019",
    address = "Florence, Italy",
    publisher = "Association for Computational Linguistics",
    url = "https://aclanthology.org/P19-1102/",
    doi = "10.18653/v1/P19-1102",
    pages = "1074--1084"
}

@inproceedings{DUC04,
  author    = {Paul Over and James Yen},
  title     = {An Introduction to {DUC-2004}: Intrinsic Evaluation of Generic News Text Summarization Systems},
  booktitle = {Proceedings of the {HLT/NAACL} 2004 Document Understanding Workshop ({DUC-2004})},
  year      = {2004},
  url       = {http://www-nlpir.nist.gov/projects/duc/pubs/2004slides/duc2004intro.pdf}
}

@misc{llama3,
      title={The Llama 3 Herd of Models}, 
      author={Aaron Grattafiori et. al.},
      year={2024},
      eprint={2407.21783},
      archivePrefix={arXiv},
      primaryClass={cs.AI},
      url={https://arxiv.org/abs/2407.21783}, 
}

@misc{rankk,
      title={Rank-K: Test-Time Reasoning for Listwise Reranking}, 
      author={Eugene Yang and Andrew Yates and Kathryn Ricci and Orion Weller and Vivek Chari and Benjamin Van Durme and Dawn Lawrie},
      year={2025},
      eprint={2505.14432},
      archivePrefix={arXiv},
      primaryClass={cs.IR},
      url={https://arxiv.org/abs/2505.14432}, 
}

@article{rbp_paper,
author = {Moffat, Alistair and Zobel, Justin},
title = {Rank-biased precision for measurement of retrieval effectiveness},
year = {2008},
issue_date = {December 2008},
publisher = {Association for Computing Machinery},
address = {New York, NY, USA},
volume = {27},
number = {1},
issn = {1046-8188},
url = {https://doi.org/10.1145/1416950.1416952},
doi = {10.1145/1416950.1416952},
journal = {ACM Trans. Inf. Syst.},
month = dec,
articleno = {2},
numpages = {27}
}


\end{document}